\documentclass[12pt]{article}
\usepackage{latexsym,amsfonts, amssymb,amsmath}
\usepackage{hyperref}
\hypersetup{CJKbookmarks= true }

\newtheorem{Theorem}{Theorem}[section]

\usepackage{enumerate}
\newtheorem{theorem}[Theorem]{Theorem}
\newtheorem{proposition}[Theorem]{Proposition}

\newtheorem{example}[Theorem]{Example}

\newcommand{\im}{\mathfrak{i}}

\def\d{{\rm diag}}
\def\k{{\rm ker}}

\usepackage{color}

\usepackage[utf8]{inputenc}
\usepackage{booktabs} 
\usepackage{tabularx}  
\usepackage{multirow} 

\begin{document}

\title{A decomposition of Hamiltonian matrices with symplectic kernels and its application to Williamson's diagonal form}
\author{Shaowu Huang$^{1} $   \footnote {Corresponding author.  \newline \indent E-mail:
shaowu2050@126.com (Shaowu Huang).}  ~~~~~ Yingxin Huang$^{2}$\\
\small $^{1}$ School of Mathematics and Finance, and Fujian Key Laboratory of Financial\\
\small Information Processing, Putian University, Putian, Fujian, People's Republic of China.\\ E-mail: shaowu2050@126.com.\\
\small $^{2}$ School of Mathematics and Finance, Putian University, \\
\small Putian, Fujian, People's Republic of China.\\ E-mail: hyxserendipity@126.com.
}

\date{}
\maketitle

\begin{abstract}
In this paper, we establish a decomposition of Hamiltonian matrices with symplectic  kernels, which extends the main result in [K. D. Ikramov,  On the Symplectic Eigenvalues of Positive Definite Matrices. Moscow Univ. Comput. Math. Cybern., 42(2018)1-4]. As an application, we present another proof of Williamson's diagonal form for positive-semidefinite matrices with symplectic kernels.

\end{abstract}

\medskip {\bf Keywords:}\  Williamson's diagonal form; Hamiltonian matrix; symplectic kernel

\medskip
{\bf 2020 Mathematics Subject Classification:}  15A23; 15A18.

\section{Introduction}
Let $A\in \mathbb{R}^{n\times m}$ be an $n\times m$ matrix with real entries. The null space and the transpose of $A$ will be denoted by $\k(A)$ and $A^T$, respectively. For $A\in \mathbb{R}^{n\times n}$, let $\sigma(A)\subseteq \mathbb{C}$ denote the spectrum (set of eigenvalues) of $A$. Set
$J_{2n}=\left(
          \begin{array}{cc}
            0 & I_n \\
            -I_n& 0 \\
          \end{array}
        \right)
$, or simply $J$ if the size is clear from context.  Here $I_n$ denotes the $n\times n$ identity.  For any $A\in \mathbb{R}^{2n\times 2n}$,  the matrix $A$ is called Hamiltonian if  $JA=(JA)^T$ and
symplectic if  $A^TJA=J$. Notice that the symplectic matrices represent the isometries of the skew-symmetric bilinear form defined by $J$, i.e., $\left\langle x,y  \right\rangle_J=x^TJy$, where $x$ and $y$ are $2n$-by-$1$ column vectors. If we partition any Hamiltonian matrix $H$ conformably with $J$, then we find
$$H=\left(
      \begin{array}{cc}
        C & L \\
        M & -C^T \\
      \end{array}
    \right),M^T=M{~\rm and~}L^T=L.
$$
Throughout this paper, $H$ will denote this block matrix.

A subspace $W\subseteq \mathbb{R}^{2n}$ is called a symplectic subspace if for any nonzero vector $x\in W$, there exists $y\in W$ such that $x^TJ_{2n}y=1$. A symplectic basis for $\mathbb{R}^{2n}$ is a collection of $2n$ vectors $\{s_1,\ldots,s_n,w_1,\ldots,w_n \}$ such that
$$\left\langle s_i,s_j \right\rangle_J = \left\langle w_i,w_j \right\rangle_J=0,\left\langle s_i,w_j \right\rangle_J = \delta_{ij},1\leq i,j \leq n,$$
where $\delta_{ij}$ is the Kronecker delta function.

Next, we introduce three types of symplectic matrices that can be used to zero selected components of a vector.
\begin{itemize}
  \item {\bf Householder symplectic matrices:}  Householder symplectic matrix $H(k,w)$ can be used to zero large portions of a vector, where
$$H(k,w)\textcolor{red}{:=}\left(
           \begin{array}{cc}
             \d(I_{k-1},P) & 0 \\
             0 &  \d(I_{k-1},P) \\
           \end{array}
         \right)
$$
and $P=I-2ww^T/w^Tw,w\in \mathbb{R}^{n-k+1}$, see \cite[Algorithm H]{loa1984}.
  \item {\bf Givens symplectic matrices:} Givens  symplectic matrix $J(k,\theta)$ can be used to zero single entry of a vector, where
$$J(k,\theta)\textcolor{red}{:=}\left(
                \begin{array}{cc}
                  C & S \\
                  -S & C \\
                \end{array}
              \right)
$$
and $C=\d(I_{k-1},\cos \theta,I_{n-k}),S=\d(0_{k-1},\sin \theta,0_{n-k})$, see \cite[Algorithm J]{loa1984}.
  \item {\bf Unipotent symplectic matrices    }: Unipotent symplectic matrix $U(n,X)$ is non-orthogonal matrix, which has form
$$U(n,X):=\left(
      \begin{array}{cc}
        I_n & X \\
        0 & I_n \\
      \end{array}
    \right),
$$where $X=\d\left(  \left(
                                                              \begin{array}{cc}
                                                                0 & \varphi \\
                                                                \varphi & 0 \\
                                                              \end{array}
                                                            \right)
 ,  0_{(n-2)\times (n-2)} \right)
$ and $\varphi\in \mathbb{R}$.
\end{itemize}

Given positive integers $n_1$, $n_2$ and $n_3$, we define
$$I_{n_1}\boxplus H(k,w) = H(k+n_1,w), I_{n_2} \boxplus  J(k,\theta)= J(k+n_2,\theta)$$
and $I_{n_3}\boxplus U(n,X)=U(n+n_3,X_1),$ where

$$X_1=\d\left(0_{n_3\times n_3} , \left(
                                                              \begin{array}{cc}
                                                                0 & \varphi \\
                                                                \varphi & 0 \\
                                                              \end{array}
                                                            \right)
 ,  0_{(n-2)\times (n-2)} \right).$$
It is clear that $H(k+n_1,w), J(k+n_2,\theta)$ and $U(n+n_3,X_1)$ are symplectic matrices.

A positive-definite(semidefinite) matrix is a symmetric matrix $A$ for which all eigenvalues are positive(nonnegative). Symplectic eigenvalues are defined for  positive-definite matrices via Williamson's diagonal form, see \cite{wil1936}. The symplectic  eigenvalues  for positive-definite matrices have many properties, including Schur-Horn type theorems (see \cite{bj2020} and our recent paper \cite{hua2023}), Cauchy's interlacing theorem (\textcolor{red}{see \cite{bj2015} and \cite{krbek2014}}), Lidskii type theorem (see \cite{jm2022}), just to name a few. For a proof of Williamson's diagonal form using a special Hamiltonian matrix decomposition, we refer to \cite[Proposition]{ikr2018}. We restate \cite[Proposition]{ikr2018} as follows:
\begin{theorem}{\rm \cite[Proposition]{ikr2018}}\label{h:th:02}
Let $A$ be a $2n\times 2n$ real symmetric positive-definite matrix. Then $JA$ can be brought by a symplectic similarity to the form
\begin{equation}\label{h:th:04}
\left(
    \begin{array}{cc}
      0 & \Pi \\
      \Sigma & 0 \\
    \end{array}
  \right),
\end{equation}
where $\Sigma$ and $\Pi$ are diagonal $n\times n$ matrices with negative and positive diagonal entries, respectively.
\end{theorem}
The authors \cite{jm2022,ss2022} presented an extension of Williamson's diagonal form to arbitrary positive-semidefinite matrices with symplectic kernels as follows:
\begin{theorem}\label{h:th:03}
For every $2n\times 2n$ real positive-semidefinite matrix $A$ with symplectic kernel, there exists a symplectic matrix $S$ such that
\begin{equation}\label{h:06}
S^TAS=\left(
          \begin{array}{cc}
            D & 0 \\
            0 & D \\
          \end{array}
        \right),
\end{equation}
where $D$ is a diagonal matrix with non-negative diagonal entries.
\end{theorem}
We shall refer to (\ref{h:06}) as a Williamson's diagonal form of $A$.

As is well known, symplectic similarity transformations preserve Hamiltonian structure. The following simple proposition will be used to prove Theorem \ref{h:th:01}.

\begin{proposition}\label{h:prop:01}
Let $H$ be a Hamiltonian matrix with symplectic kernel and $S$ be a symplectic matrix. Then $SHS^{-1}$ is a Hamiltonian matrix with symplectic kernel. Moreover, if  $C,L$ and $M$ are $n\times n$ diagonal matrices, then $H_k=\left(
                                                                      \begin{array}{cc}
                                                                        c_k & l_k \\
                                                                        m_k & -c_k \\
                                                                      \end{array}
                                                                    \right),k=1,2,\ldots,n
$ are Hamiltonian matrices with symplectic kernels.

\end{proposition}
\it Proof. \rm Let $\dim(\k(H))=2k$ and $\{s_1,\ldots,s_k,w_1,\ldots,w_k\}$ be a symplectic basis of $\k(H)$. Since
$$\dim(\k(SHS^{-1}))=2n-{\rm rank}(SHS^{-1})=2n-{\rm rank}(H)=\dim(\k(H)),$$
we have
$$\{Ss_1,\ldots,Ss_k,Sw_1,\ldots,Sw_k\}$$
is a symplectic basis of $\k(SHS^{-1})$.  Thus, by \cite[Lemma 2.1]{ss2022}, $\k(SHS^{-1})$ is a symplectic subspace.

For any nonzero vector  $(c~z)^T\in \k(H_k)$, we get $x=\left( x_1,x_2,\ldots,x_{2n}   \right)^T\in \k(H)$, where
$$x_i=\left\{
    \begin{array}{ll}
      0, & \hbox{$i\ne k,n+k$;} \\
      c, & \hbox{$i=k$;} \\
      z, & \hbox{$i=n+k$.}
    \end{array}
  \right.
$$
Since the kernel of $H$ is  a symplectic subspace of $\mathbb{R}^{2n}$,  there exists a vector $y=(y_1,y_2,\ldots,y_{2n})^T\in \k(H)$ such that $x^TJy=1$.

Hence, we obtain $H_k(y_k~y_{n+k})^T=0$ and
$$(c~z)^TJ\left(
                    \begin{array}{c}
                      y_k \\
                      y_{n+k} \\
                    \end{array}
                  \right)=x^TJy=1,
$$
which implies the kernel of  $H_k$ is a symplectic subspace. $\square$

\begin{example}\label{h:ex:01}
Let $n\ge 2$, $x,y\in \mathbb{R}^{2n}$ with $\left\langle x,y \right\rangle_J \ne 0$. Then there exists a symplectic matrix $S$ such that
$$S\left(
     \begin{array}{cc}
       x & y \\
     \end{array}
   \right)
=\left(
            \begin{array}{c}
              W_1 \\
              W_2 \\
            \end{array}
          \right)
 ,W_1=\left(
           \begin{array}{cc}
             \alpha & \beta \\
             0 & 0 \\
              \vdots & \vdots \\
             0 & 0\\
           \end{array}
         \right)\in \mathbb{R}^{n\times 2}{ ~and~}W_2=\left(
           \begin{array}{cc}
            0 & \gamma \\
             0 & 0 \\
              \vdots & \vdots \\
             0 & 0\\
           \end{array}
         \right)\in \mathbb{R}^{n\times 2},
$$
where $\alpha\ne 0, \gamma\ne 0$ and $S$ is a product of Householder symplectic matrices, Givens symplectic matrices and a unipotent symplectic matrix. The case $n=1$ is trivial.
\end{example}

In fact, observe that

\begin{eqnarray}\nonumber
\left(
     \begin{array}{cc}
       x & y \\
     \end{array}
   \right) & \rightarrow \overset{G_1}{ \cdots } \rightarrow & \left(
                                                                     \begin{array}{cc}
                                                                       x_1^{(1)} & y_1^{(1)}\\
                                                                       0 & y_2^{(1)} \\
                                                                     \end{array}
                                                                   \right)   \overset{\textcolor{red}{H_1(k,w)}}{ \rightarrow }
                                                                   \left(
                                                                     \begin{array}{cc}
                                                                       x_1^{(2)} & y_1^{(2)}\\
                                                                       0 & y_2^{(2)} \\
                                                                     \end{array}
                                                                   \right) \\
    &\rightarrow \overset{G_2}{ \cdots } \rightarrow  &  \left(
   \begin{array}{cc}
     x_1^{(2)} & y_1^{(3)} \\
     0 & y_2^{(3)} \\
   \end{array}
 \right) \overset{\textcolor{red}{H_2(k,w)}}{ \rightarrow }    \left(
   \begin{array}{cc}
     x_1^{(2)} & y_1^{(4)} \\
     0 & y_2^{(3)} \\
   \end{array}
 \right)\label{h:07}
\end{eqnarray}
where $\alpha\ne 0,\gamma\ne 0$,
$$x_1^{(2)}=\left( \alpha,0,\ldots,0    \right)^T, y_2^{(3)}=\left( \gamma,0,\ldots,0    \right)^T,y_1^{(4)}=\left(\beta,\delta,0,\ldots,0    \right)^T$$
and
\begin{eqnarray*}
H_1(k,w) &=& H(1,w_1), H_2(k,w)=H(2,w_2),\\
  G_1 &=& J(1,\theta_1) J(2,\theta_2) \cdots J(n,\theta_n), G_2 =  J(2,\theta_2') J(3,\theta_3') \cdots J(n,\theta_n').
\end{eqnarray*}
We now use  $\varphi=-\frac{\delta}{\gamma}$ and (\ref{h:07}) to derive
$$U(n,X)\left(
   \begin{array}{cc}
     x_1^{(2)} & y_1^{(4)} \\
     0 & y_2^{(3)} \\
   \end{array}
 \right)=  \left(
            \begin{array}{c}
              W_1 \\
              W_2 \\
            \end{array}
          \right). $$

\begin{example}\label{h:ex:02}
Let $H=\left(
         \begin{array}{cc}
           s & w \\
           \textcolor{red}{v} & -s \\
         \end{array}
       \right)\in \mathbb{R}^{2\times 2}
$ with symplectic kernel and its eigenvalues be $\lambda=\pm \im a,a\in \mathbb{R}$. Then there exists a  symplectic matrix $S$ such that
$$SHS^{-1}=\left(
            \begin{array}{cc}
              0 & p \\
              q & 0 \\
            \end{array}
          \right),
$$
where $pq\leq 0$ and $p=0$ if and only if $q=0$.

\end{example}

We start with case 1, that is we assume that $w+v=0$.  If we set $S=\frac{\sqrt{2}}{2}\left(
                               \begin{array}{cc}
                                 1 & 1 \\
                                 -1 & 1 \\
                               \end{array}
                             \right)
$, it is easy to show that
$$SHS^{-1}=\frac{1}{2}\left(
                 \begin{array}{cc}
                   0 & w-v-2s \\
                   v-w-2s & 0 \\
                 \end{array}
               \right)=\left(
                         \begin{array}{cc}
                           0 & -(s-w) \\
                           -(s+w) & 0 \\
                         \end{array}
                       \right).
$$

Now we consider case 2, namely we assume that $w+v\ne 0$. If we take $S=\left(
                                                                                                    \begin{array}{cc}
                                                                                                      \cos \theta& \sin \theta\\
                                                                                                      -\sin \theta& \cos \theta\\
                                                                                                    \end{array}
                                                                                                  \right)
$ with $\tan 2\theta=\frac{-2s}{w+\textcolor{red}{v}}$ , then we have
\begin{eqnarray*}
SHS^{-1} &=& \left(
             \begin{array}{cc}
               \cos 2\theta \cdot s + \sin 2\theta (w+\textcolor{red}{v})/2 &  p_1\\
             n_1   & -\cos 2\theta \cdot s - \sin 2\theta (w+\textcolor{red}{v})/2 \\
             \end{array}
           \right) \\
   &=& \left(
                     \begin{array}{cc}
                       0 & p_1 \\
                       \textcolor{red}{q_1} & 0 \\
                     \end{array}
                   \right).
\end{eqnarray*}

Therefore, in either case, there exists a symplectic matrix $S$ such that
$$SHS^{-1}=\left(
            \begin{array}{cc}
              0 & p \\
              \textcolor{red}{q} & 0 \\
            \end{array}
          \right).
$$
Since $H$ has eigenvalues $\lambda=\pm \im a,a\in \mathbb{R}$, we have $pn\leq 0$. Moreover, if $p=0$, then $0\in \sigma(SHS^{-1})$. By Proposition \ref{h:prop:01}, we have  $SHS^{-1}(x~y)=0,$ where $\{x,y\}$ is a symplectic basis of $\k\left(SHS^{-1}\right)$. Since $SHS^{-1}(x~y)=0
$, we have $H=0$, and then \textcolor{red}{$q=0$}. Similarly, it follows from $q=0$ that $p=0$.

In this article, we study  Hamiltonian matrices decompositions with symplectic kernels beyond those mentioned above, see Theorem \ref{h:th:01} and Theorem \ref{h:th:05}. We use three types of symplectic matrices as our tools to explore the existence of a decomposition for  Hamiltonian matrices in Section 2. Moreover, we prove some general results for the Hamiltonian matrix $JA$ with symplectic kernel and establish a strategy to present another proof of Theorem \ref{h:th:03}  in Section 3.

\section{A decomposition of  Hamiltonian matrices with symplectic kernels}

Let us recall that  $JA$ admits  the desired decomposition (\ref{h:th:04}) for any positive-definite matrix $A$.  However, in general, Theorem \ref{h:th:02} does not hold for all Hamiltonian matrices which have only pure imaginary eigenvalues and don't  have an eigenvector $z=x+\im y$ for which $\left\langle  x,y \right\rangle_J=0$, where $x\ne 0,y\ne 0$. For example, let $H_2=-J\in \mathbb{R}^{2\times 2}$ be a Hamiltonian matrix. Clearly, $\pm \im $ are its  eigenvalues and let $H_2(c+\im m)=\im a (c+\im m)$, where $a= \pm 1, c\ne 0, m\ne0$. It is easy to verify that $H_2c=-am,H_2m=ac$ and $m=-aH_2c$.  Therefore, in this example,  we have $c^TJm=c^TJ\left( -a(-J)c \right)=-ac^Tc \ne 0$.But
\begin{eqnarray*}
H_2' &=& SH_2S^{-1}=\left(
                    \begin{array}{cc}
                      a & b \\
                      \textcolor{red}{f} & d \\
                    \end{array}
                  \right)\left(
                           \begin{array}{cc}
                             0 & -1 \\
                             1 & 0\\
                           \end{array}
                         \right)\left(
                                  \begin{array}{cc}
                                    d & -b \\
                                    \textcolor{red}{-f} & a \\
                                  \end{array}
                                \right) \\
   &=& \left(
                                          \begin{array}{cc}
                                            bd+a\textcolor{red}{f} & -(b^2+a^2) \\
                                            d^2 +\textcolor{red}{f^2} & -bd -a\textcolor{red}{f} \\
                                          \end{array}
                                        \right)
\end{eqnarray*}
for any symplectic matrix $S=\left(
                    \begin{array}{cc}
                      a & b \\
                      \textcolor{red}{f} & d \\
                    \end{array}
                  \right)$, which implies $H_2$ does not have  the decomposition (\ref{h:th:04}).

In this section,  we  prove an analogous result to  Theorem \ref{h:th:02}. For any Hamiltonian matrix $H$ with symplectic kernel,  $H$ has a decomposition of an analogous form  (\ref{h:th:04}).

\begin{theorem}\label{h:th:01}
Let $H\in \mathbb{R}^{2n\times 2n}$ be a Hamiltonian matrix with symplectic kernel and all its eigenvalues be 0 or purely imaginary over $\mathbb{C}$.
For any nonzero eigenvalue $\lambda$ and $c,m\in \mathbb{R}^{2n}$, if $\lambda$ does not have an eigenvector $z=c+\im m$ such that $\left\langle c,m  \right\rangle_J=0$, then there exists a symplectic matrix $S$ such that
$$H=S^{-1}\left(
        \begin{array}{cc}
          0 & P \\
          N & 0 \\
        \end{array}
      \right)S,
$$
where $S$ is a product of Householder symplectic matrices, Givens symplectic matrices and  unipotent symplectic matrices, $P=\operatorname{diag}(p_{11},\dots,p_{nn})$ and $N=\operatorname{diag}(q_{11},\dots,q_{nn})$ are $n\times n$ diagonal  matrices with
\begin{equation}\label{h:04}
  \left\{
   \begin{array}{ll}
     q_{ii}=0, & if ~\hbox{$p_{ii}=0$;} \\
     q_{ii}p_{ii}<0, & if ~\hbox{$p_{ii}\ne 0$.}
   \end{array}
 \right.
\end{equation}

\end{theorem}

\it Proof. \rm We claim that $H$ admits the following decomposition
\begin{equation}\label{h:03}
S_AHS_A^{-1}=\left(
                 \begin{array}{cc}
                   D^{(1)} & D^{(2)} \\
                   D^{(3)} & -D^{(1)} \\
                 \end{array}
               \right),
\end{equation}
where $S_A$ is a product of Householder symplectic matrices, Givens symplectic matrices and  unipotent symplectic matrices,  $D^{(l)}=\operatorname{diag}\big(d_{11}^{(l)},\dots,d_{nn}^{(l)}\big)
$, $l=1,2,3$ are diagonal matrices and
$$\left\{
    \begin{array}{ll}
      d^{(l)}_{ii}=0~(l=1,3), & \hbox{\textcolor{red}{\text{if }}$d^{(2)}_{ii}=0$;} \\
      d^{(3)}_{ii}\ne 0, & \hbox{\textcolor{red}{\text{if }}$d^{(2)}_{ii}\ne 0$.}
    \end{array}
  \right.
$$

Now we prove this claim by induction on $n$. Suppose  $H=\left(
                                              \begin{array}{cc}
                                                a & b \\
                                                c & -a \\
                                              \end{array}
                                            \right)
$, we distinguish two cases.

  Case 1:  $b=0$. Since all eigenvalues of $H$ are 0 or purely imaginary over $\mathbb{C}$, we find $a^2+bc\leq 0$, and then $a=0$. In this case, $0\in \sigma(H)$. Let $\{x,y\}$ be a symplectic basis of $\k(H)$. From the fact $H(x~y)=0
$ and $x,y$ are linearly independent, it follows that $H=0$.

  Case 2:  $b\ne 0$. Assume, to the contrary, $c=0$. This implies $a=0$. By the same arguments as in the proof of Case 1, we find $H=0$, leading to a contradiction.

The required statement for $n=1$ follows from the above two cases. Assume now that $n\geq 2$ and that the statement holds for $n-1$. We now show that there exists a symplectic matrix $S_1$ such that
\begin{equation}\label{h:08}
S_1HS_1^{-1}=\left(
                 \begin{array}{cccc}
                   -\frac{\beta}{\alpha}l & 0 &  \frac{\alpha^2 + \beta^2}{\alpha \gamma}l & 0 \\
                   0 & C_1 & 0 & L_1 \\
                   -\frac{\gamma}{\alpha}l & 0 & \frac{\beta}{\alpha}l & 0 \\
                   0 & M_1 & 0 & -C_1^T \\
                 \end{array}
               \right),
\end{equation}
where $\alpha\ne 0, \gamma\ne 0$ and $\lambda=\im l,l\in \mathbb{R}$. We may consider two cases depending on the  values of $l$.

Case 1: $l=0$. Let $\dim(\k (H))=2k$ and   $\{s_1,\ldots,s_k,w_1,\ldots,w_k\}$ be a symplectic basis of $\k (H)$.  By $s_1^T Jw_1=1$ and Example \ref{h:ex:01}, there exists $S_{11}$ such that
$$S_{11}(s_1~w_1)=\left(
                                         \begin{array}{c}
                                           W_1 \\
                                           W_2 \\
                                         \end{array}
                                       \right) ,W_1=\left(
           \begin{array}{cc}
             \alpha & \beta \\
             0 & 0 \\
              \vdots & \vdots \\
             0 & 0\\
           \end{array}
         \right)\in \mathbb{R}^{n\times 2}{\rm ~and~}W_2=\left(
           \begin{array}{cc}
            0 & \gamma \\
             0 & 0 \\
              \vdots & \vdots \\
             0 & 0\\
           \end{array}
         \right)\in \mathbb{R}^{n\times 2},
$$
where $\alpha\ne 0, \gamma\ne 0$.

Partition $S_{11}HS_{11}^{-1}$ as $S_{11}HS_{11}^{-1}=(h_1, \ldots ,h_{2n})$. It follows from the conditions $S_{11}HS_{11}^{-1}S_{11}s_1=S_{11}HS_{11}^{-1}S_{11}w_1=0$ that $h_1=h_{n+1}=0$. Since $S_{11}HS_{11}^{-1}$  is  Hamiltonian, the matrix $S_{11}HS_{11}^{-1}$ is of the desired form (\ref{h:08}).

Case 2: $l\ne 0$.  Let $H(c+\im m)=\im l(c+\im m),c,m\in \mathbb{R}^{2n\times 1}$. We have $\left\langle c,m  \right\rangle_J\ne 0$ and $Hc=-lm,Hm=lc$. Hence, by Example \ref{h:ex:01}, there exists a symplectic matrix $S_{12}$ such that
\begin{equation}\label{h:02}
S_{12}HS_{12}^{-1}S_{12}c=-lS_{12}m,S_{12}HS_{12}^{-1}S_{12}m=lS_{12}c,
\end{equation}
where $\alpha'\ne 0, \gamma'\ne 0$ and
$$S_{12}(c~m)=\left(
                                         \begin{array}{c}
                                           W_1' \\
                                           W_2' \\
                                         \end{array}
                                       \right) ,W_1'=\left(
           \begin{array}{cc}
             \alpha' & \beta' \\
             0 & 0 \\
              \vdots & \vdots \\
             0 & 0\\
           \end{array}
         \right)\in \mathbb{R}^{n\times 2},W_2'=\left(
           \begin{array}{cc}
            0 & \gamma' \\
             0 & 0 \\
              \vdots & \vdots \\
             0 & 0\\
           \end{array}
         \right)\in \mathbb{R}^{n\times 2}.
$$

Partition $S_{12}HS_{12}^{-1}$ as $S_{12}HS_{12}^{-1}=(h'_1, \ldots ,h'_{2n})$.
By (\ref{h:02}), we have
\begin{eqnarray*}
  \alpha' h'_1 &=& -l\left(
                    \begin{array}{cccccccc}
                      \beta' & 0 & \cdots & 0 & \gamma' & 0 & \cdots & 0 \\
                    \end{array}
                  \right)^T
 \\
  h'_1&=&-\frac{l}{\alpha'}\left(
                    \begin{array}{cccccccc}
                      \beta' & 0 & \cdots & 0 & \gamma' & 0 & \cdots & 0 \\
                    \end{array}
                  \right)^T
\end{eqnarray*}
and
\begin{eqnarray*}
\beta' h'_1 +\gamma' h'_{n+1} &=& l\left(
                                 \begin{array}{cccc}
                                   \alpha' & 0 & \cdots & 0 \\
                                 \end{array}
                               \right)^T
 \\
  h'_{n+1} &=& \left(
                \begin{array}{cccccccc}
                 \frac{\alpha'^2 + \beta'^2}{\alpha' \gamma'}l  & 0 & 0\cdots & 0 &  \frac{\beta'}{\alpha'}l  & 0 & \cdots & 0 \\
                \end{array}
              \right)^T.
\end{eqnarray*}
Using the above results and $S_{12}HS_{12}^{-1}$  is  Hamiltonian, we obtain the desired form (\ref{h:08}).

Setting
$$H_1=\left(
        \begin{array}{cc}
          C_1 & L_1 \\
          M_1 & -C_1^T \\
        \end{array}
      \right),H_2=\left(
                    \begin{array}{cc}
                       -\frac{\beta}{\alpha}l & \frac{\alpha^2 + \beta^2}{\alpha \gamma}l\\
                      -\frac{\gamma}{\alpha}l & \frac{\beta}{\alpha}l  \\
                    \end{array}
                  \right),
$$
we see that $H_1$ is a Hamiltonian matrix.   It remains to show $H_1$ satisfies the conditions in this theorem. From the block form of $S_1HS_1^{-1}$, we obtain $\sigma(H)=\sigma(H_1)\cup \sigma(H_2)$, and then all eigenvalues of  $H_1$  are 0 or purely imaginary over $\mathbb{C}$. For  any $Z_1\in \ker (H_1)$, partition
$$Z_1=\left(
        \begin{array}{cc}
          z_{11} & z_{21} \\
        \end{array}
      \right)
^T,Z=\left(
       \begin{array}{cccc}
         0 & z_{11} & 0 & z_{21} \\
       \end{array}
     \right)^T.$$
Clearly, $Z\in \ker(S_1HS_1^{-1})$. By Proposition \ref{h:prop:01}, we have $Z^TJY=1$ for some $Y=(*~y_{11}~*~y_{21})^T \in \ker(S_1HS_1^{-1})$. So
$$Z_1^TJ_{2(n-1)}\left(
          \begin{array}{c}
            y_{11} \\
            y_{21} \\
          \end{array}
        \right)
=Z^TJ_{2n}Y=1,$$
which means that the kernel of $H_1$ is a  symplectic subspace.

On the other hand, let $\left(
                          \begin{array}{cc}
                            c_1 & c_2 \\
                          \end{array}
                        \right)^T+\im \left(
                                      \begin{array}{cc}
                                        m_1 & m_2\\
                                      \end{array}
                                    \right)^T
$ be any  eigenvector of $H_1$ with $\left(
                          \begin{array}{cc}
                            c_1 & c_2 \\
                          \end{array}
                        \right)\ne 0$ and $ \left(
                                      \begin{array}{cc}
                                        m_1 & m_2\\
                                      \end{array}
                                    \right)\ne 0$. It follows that  $\zeta$ and $S_1^{-1}\zeta$ are the eigenvectors of $S_1HS_1^{-1}$ and $H$, respectively, where
$$\zeta=\left(
                       \begin{array}{cccc}
                         0 & c_1 & 0 & c_2\\
                       \end{array}
                     \right) +\im \left(
                                    \begin{array}{cccc}
                                      0 & m_1 & 0& m_2 \\
                                    \end{array}
                                  \right)\triangleq \zeta_1 + \im \zeta_2.$$
With the help of $\left\langle S_1^{-1}\zeta_1,S_1^{-1}\zeta_2  \right\rangle_J\ne 0$, it is readily checked that
$$\left(
    \begin{array}{cc}
      c_1 & c_2 \\
    \end{array}
  \right)J\left(
            \begin{array}{c}
              m_1 \\
              m_2 \\
            \end{array}
          \right)= \left( S_1^{-1}\zeta_1 \right)^T J \left(  S_1^{-1}\zeta_2 \right) \ne 0.
$$

The induction hypothesis implies that there exist symplectic matrices $H(k_i,w_i),$ $J(k_j,\theta_j)$ and $U(n_l,X_l)$ such that
$$S_2H_1S_2^{-1}=\left(
                 \begin{array}{cc}
                   \tilde{D}^{(1)} & \tilde{D}^{(2)} \\
                   \tilde{D}^{(3)} & -\tilde{D}^{(1)} \\
                 \end{array}
               \right)
$$
where $1\leq i \leq s_1,1\leq j\leq s_2$, $1\leq l\leq s_3 $ and $S_2$ is a product of the above three types of symplectic matrices.

Let $S_A$ be a product of the following three types of symplectic matrices
$$1\boxplus H(k_i,w_i),1\boxplus J(k_j,\theta_j){\rm~and~}1\boxplus U(n_l,X_l).$$
We get the required result (\ref{h:03}).

In view of Example \ref{h:ex:02}, there exists a symplectic matrix  $S_3$ such that
$$S_3S_AHS_A^{-1}S_3^{-1}=\left(
                              \begin{array}{cc}
                                0 & P \\
                                N & 0 \\
                              \end{array}
                            \right),
$$
where $S_3$ is a product of Givens symplectic matrices, $P$ and $N$ are diagonal matrices and satisfy (\ref{h:04}). $\square$

\section{A decomposition of $JA$ with symplectic kernel}
This section introduces a decomposition of   $JA$ with symplectic kernel. We first give some results about $JA$.

\begin{proposition}\label{h:prop:02}
Let $A$ be a $2n\times 2n$  real symmetric positive-semidefinite matrix. Then
\begin{enumerate}[\rm (a)]
  \item all eigenvalues of $JA$ are $0$ or $\im a (a\in \mathbb{R},a\ne 0)$ over $\mathbb{C}^{2n}$;
  \item  if $z=x+\im y$ is any complex eigenvector of $JA$ corresponding to the eigenvalue $\lambda=\im a, 0\ne a\in \mathbb{R}$ over $\mathbb{C}^{2n}$, then $\left< x,y \right>_J\ne 0$.
\end{enumerate}

\end{proposition}

\it Proof. \rm (a) Let $JAz=\lambda z,z=x+\im y,x,y\in \mathbb{R}^{2n}$. Then
\begin{equation}\label{h:01}
z^* Az =- \lambda z^*Jz.
\end{equation}
Here $z^*$ is the complex conjugate and transpose of complex $z$.

On the other hand, $z^*Jz=(z^*Jz)^T=-z^TJ\overline{z}=-\overline{(z^*Jz)}$. So we have $z^*Jz=\im k,k\in \mathbb{R}$. If $k\ne 0$, then, by (\ref{h:01}), we have $-\lambda (\im k)=z^*Az\in \mathbb{R}$,  and then $\lambda=\im a$ for some $a \in \mathbb{R}$ (where $a\neq 0$ in general); If $k=0$,  then
 $$z^*Az=0\Rightarrow A^{1/2}z=0\Rightarrow Az=0.$$
Hence, $\lambda=0$.

(b) Let $\lambda=\im a\ne 0$,$z=x+\im y$ and $JAz=\lambda z$.  We must have $x,y\ne 0$ and claim that $x^T Ax + y^T Ay > 0$, because otherwise,  if $x^T Ax + y^T Ay = 0$, then  $x^T Ax = y^T Ay = 0$ implies $Ax=Ay=0 $. It follows that $\lambda z=0$, and then $\lambda=0$, which contradicts $\lambda\ne 0$.

On the other hand,
\begin{eqnarray*}
  2\left\langle x,y\right>_J &=&- \left( y^TJx +y^TJx \right)= \frac{1}{a}y^TAy +\frac{1}{a}(JAx)^TJx \\
    &=&\frac{1}{a}y^TAy - \frac{1}{a}x^TAJJx = \frac{1}{a} \left(  x^T Ax + y^T Ay  \right)\\ & \ne & 0,
\end{eqnarray*}
which completes  the proof. $\square$

\bf Remark. \rm The first part of Proposition \ref{h:prop:02} was proved by using  Williamson's diagonal form, which has been discussed in \cite[Proposition 2.1]{km2026}.

Next we show how Theorem \ref{h:th:02} can naturally be extended to positive-semidefinite matrices with symplectic  kernels.
\begin{theorem}\label{h:th:05}
Let $A\in \mathbb{R}^{2n\times 2n}$ be a symmetric positive-semidefinite matrix with symplecitc kernel. Then
\begin{equation}\label{h:05}
  JA=S\left(
        \begin{array}{cc}
          0 & P \\
         N & 0 \\
        \end{array}
      \right)S^{-1},
\end{equation}
where $S$ is a symplectic matrix, $P=(p_{ij})$ and $N=(n_{ij})$ are $n\times n$ diagonal  matrices with non-negative and non-positive diagonal entries, respectively. Moreover, $p_{ii}> 0$ if and only if $n_{ii}< 0$.
\end{theorem}
\it Proof. \rm
Using Theorem \ref{h:th:01} and Proposition \ref{h:prop:02}, we see that there exists a symplectic matrix $S$ such that
\begin{eqnarray*}
S^{-1}JAS &=& \left(
        \begin{array}{cc}
          0 & P \\
         N & 0 \\
        \end{array}
      \right) ,
\end{eqnarray*}
where $P$ and $N$ can be obtained with the procedure under the properties in (\ref{h:04}). Note that $S^{-1}JS^{-T}=J$ and it is apparent that
$$
  S^TAS = \left(
        \begin{array}{cc}
          -N & 0 \\
         0 & P \\
        \end{array}
      \right).$$
Hence, $-N$ and $P$ are positive-semidefinite. $\square$

\bf Another proof of Theorem  \ref{h:th:03}. \rm In view of Theorem \ref{h:th:05}, there exists a symplectic matrix \textcolor{red}{$S_1$ }and diagonal matrices $P$ and $N$ defined in (\ref{h:05}) such that
$$  JA=\textcolor{red}{S_1^{-1}}\left(
        \begin{array}{cc}
          0 & P \\
         N & 0 \\
        \end{array}
      \right)\textcolor{red}{S_1},$$
which implies, just as in the proof of Theorem  \ref{h:th:05}, that
$$S_1^TAS_1 = \left(
        \begin{array}{cc}
          -N & 0 \\
         0 & P \\
        \end{array}
      \right).$$

Let us define the symplectic matrix $\Phi$ as follows:
$$\Phi=\d \left(\phi_1, \ldots, \phi_n, \phi_1^{-1}, \ldots, \phi_n^{-1} \right) {\rm ~and~} \phi_i=\left\{
           \begin{array}{ll}
             1, & \hbox{\textcolor{red}{\text{if }}$n_{ii}=0$;} \\
             \left( \frac{p_{ii}}{-n_{ii}} \right)^{\frac{1}{4}}, & \hbox{\textcolor{red}{\text{if }}$n_{ii}\ne 0$.}
           \end{array}
         \right. $$
Since the product of two symplectic matrices is symplectic, $S_1\Phi$ is symplectic.

Therefore,
$$\Phi S_1^TAS_1\Phi=(S_1\Phi)^T A (S_1\Phi)=\left(
                   \begin{array}{cc}
                     D & 0 \\
                     0 & D \\
                   \end{array}
                 \right),
$$
where $D=\d(d_1,\ldots,d_n)$ and $d_i\geq 0,i=1,\ldots,n$.

\section*{Acknowledgement}
The work was supported by the Fujian Provincial Natural Science Foundation of China [Nos.2024J01874, 2026J0011063].


\begin{thebibliography}{99}
\bibitem{bj2015} R. Bhatia, T. Jain, On the symplectic eigenvalues of positive definite matrices, J. Math. Phys., 56(2015)112201.
\bibitem{bj2020}R. Bhatia, T. Jain, A Schur-Horn theorem for symplectic eigenvalues, Linear Algebra Appl.,  599(2020)133-139.
\bibitem{hua2023}S. Huang, A new version of Schur-Horn type theorem, Linear and Multilinear Algebra, 71(1)(2023)41-46.
\bibitem{loa1984} C. Van Loan, A symplectic method for approximating all the eigenvalues of a Hamiltonian matrix, Linear Algebra Appl., 61(1984)233-251.
\bibitem{ikr2018}K. D. Ikramov,  On the Symplectic Eigenvalues of Positive Definite Matrices. Moscow Univ. Comput. Math. Cybern., 42(2018)1-4.
\bibitem{jm2022}T. Jain, H.K. Mishra, Derivatives of symplectic eigenvalues and a Lidskii type theorem. Canad. J. Math., 74(2022)457-485.
\bibitem{km2026} R. R. Kamat, H. K. Mishra, Simultaneous symplectic spectral decomposition of positive semidefinite matrices, Linear Algebra Appl.,  737(2026)298-307.
\bibitem{ss2022} N. T. Son, T. Stykel, Symplectic eigenvalues of positive-semidefinite matrices and the trace minimization theorem, Electron. J. Linear Algebra, 38(2022)607-616.


\bibitem{wil1936}J. Williamson, On the algebraic problem concerning the normal forms of linear dynamical systems, Am. J. Math., 58(1)(1936)141-163.
\bibitem{krbek2014} M. Krbek, T. Tyc, J. Vlach, Inequalities for quantum marginal problems with continuous variables, J. Math. Phys., 55(2014) 062201.



\end{thebibliography}
\end{document}